\documentclass[aps,twocolumn,showpacs,amsmath,amssymb,floatfix,nofootinbib]{revtex4}

\usepackage{graphicx}
\usepackage{dcolumn}
\usepackage{bm}

\begin{document}
\title{Apparent Hubble acceleration from\\
large-scale electroweak domain structure\\}

\author{Tommy Anderberg}
\email{Tommy.Anderberg@simplicial.net}
\noaffiliation

\date{\today}

\begin{abstract}
The observed luminosity deficit of Type Ia supernovae (SNe Ia) at
high redshift $z$ can be explained by partial conversion to weak
vector bosons of photons crossing large-scale electroweak domain
boundaries, making Hubble acceleration only apparent and eliminating
the need for a cosmological constant $\Lambda > 0$.
\end{abstract}


\maketitle


\section{Introduction}

After the initial surprise caused by the announcement of a
luminosity deficit from high redshift $z$ Type Ia supernovae (SNe
Ia) \cite{riess1998}\cite{perlmutter1998}\cite{perlmutter1999}, the
consensus quickly emerged that we are witnessing an accelerating
Hubble expansion, implying a cosmological constant $\Lambda > 0$.
This has forced a profound change in our view of the large scale
structure and composition of the universe, leading us from a
preferred model with negative curvature and density parameter
$\Omega \approx 0.3$ dominated by (mostly dark) matter to a flat
geometry with $\Omega \approx 1$ dominated by the dark energy term
$\Omega_\Lambda \approx 0.7$ ($\Lambda$CDM). Most importantly from
the point of view of fundamental physics, it has also left us with
the massive embarrassment of a {\em finite} $\Lambda$ 120 orders of
magnitude below the Planck scale, where it's customarily argued on
dimensional grounds that the zero point energy of quantum fields
coupled to gravity should show up, absent some fundamental principle
forcing it to vanish exactly.

These far-reaching implications have motivated many studies of
alternative explanations for the dimming of high-$z$ SNe Ia, from
variations in intrinsic luminosity (chemical abundances, stellar
populations) through modified light propagation (gravitational
lensing, gray dust) to systematic observational errors (selection
bias) \cite{filippenko2003}\cite{riess2004}. But to date, all
suggested mechanisms have proved incapable of producing effects of
the needed size, -0.25 magnitudes at $z \sim 0.5$ (where the
luminosity deficit has its greatest leverage on $\Lambda$), i.e. a
ratio between observed luminosity $\ell$ and expected luminosity
$\ell_E$
\begin{eqnarray}
\ell / \ell_E \approx 100^{-0.25/5} \approx 0.79
\end{eqnarray}
The case for dark energy rests squarely on this number. In spite of
common claims to the contrary, dimming of SNe Ia (and now of gamma
ray bursts \cite{schaefer2006}) remains the only direct evidence of
accelerated expansion \cite{blanchard2003}. Even disregarding the
possibility of significant systematic errors in distance measures
\cite{bonanos2006}\cite{wang2006}, the oft-quoted consistency of
$\Lambda$CDM with other observations is only a necessary condition
for its validity, not a sufficient one. The true power of this
condition is its ability to falsify the framework: find an
explanation for one data set leaving insufficient room for
accelerated expansion to fit the others and the entire framework is
invalidated.

The main objective of this paper is to show how the standard
electroweak and big bang models can lead to the observed $\ell /
\ell_E$ without any Hubble acceleration actually taking place, and
how they are in fact constrained by observation to do so in a way
which leaves little room for $\Lambda > 0$.

\section{Photon transformations}

\begin{figure}
\includegraphics[scale=0.55]{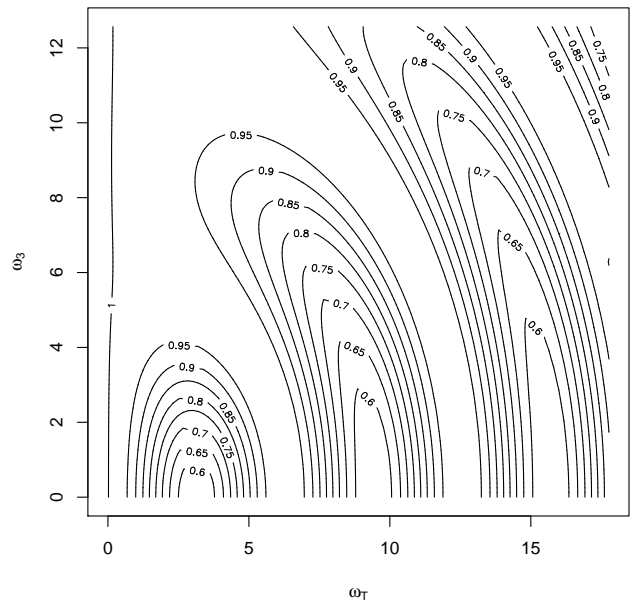}
\caption{\label{fig:coeff_contour}Residual luminosity $\ell({\bm
\omega})$ for $\omega_\bot \in [0, \sqrt{32}\pi]$, $\omega_3 \in [0,
4\pi]$ (contour plot).}
\end{figure}

\begin{figure}
\includegraphics[scale=0.55]{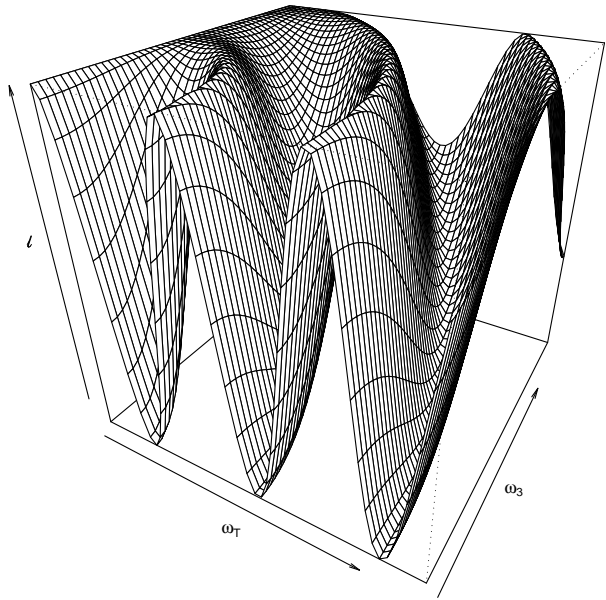}
\caption{\label{fig:coeff_persp}Residual luminosity $\ell({\bm
\omega})$ for $\omega_\bot \in [0, \sqrt{32}\pi]$, $\omega_3 \in [0,
4\pi]$ (perspective plot).}
\end{figure}

In the standard SU(2)$\times$U(1) model of electroweak interactions
\cite{glashow1961}\cite{weinberg1967}\cite{salam1968}, the U(1)
gauge field $B_\mu(x)$ and the three SU(2) gauge fields $W^j_\mu(x)$
are collected in the matrix (the field-dependent part of the
covariant derivative)
\begin{eqnarray}\label{defm}
{\bm M}_\mu(x) \equiv \left[
\begin{array}{lr}
g' B_\mu(x) + g W^3_\mu(x) & g W^1_\mu(x) - i g W^2_\mu(x)\\
g W^1_\mu(x) + i g W^2_\mu(x) & g' B_\mu(x) - g W^3_\mu(x)
\end{array}
\right]
\end{eqnarray}
($g$ = SU(2) coupling constant; $g'$ = U(1) coupling constant)
acting on weak isospin doublets. Given an arbitrary ${\bm
M}_\mu(x)$, we can decompose it into individual gauge fields using
\begin{eqnarray}\label{defb}
B_\mu(x) &=& \frac{1}{2g'} {\rm Tr}\left[\tau_0 \thinspace {\bm M}_\mu(x)\right] \\
W^1_\mu(x) &=& \frac{1}{2g} {\rm Tr}\left[\tau_1 \thinspace {\bm M}_\mu(x)\right] \\
W^2_\mu(x) &=& \frac{i}{2g}{\rm Tr}\left[\tau_2 \thinspace {\bm M}_\mu(x)\right] \\
W^3_\mu(x) &=& \frac{1}{2g} {\rm Tr}\left[\tau_3 \thinspace {\bm
M}_\mu(x)\right] \label{defw3}
\end{eqnarray}
where all traces are understood to be over weak isospin space only,
$\tau_0 \equiv {\bm 1}_2$ and $\tau_1$, $\tau_2$, $\tau_3$ are the
Pauli matrices
\begin{eqnarray}
\tau_1 \equiv \left[
\begin{array}{lr}
0 & 1\\
1 & 0
\end{array}
\right]\quad \tau_2 \equiv \left[
\begin{array}{lr}
0 & -i\\
i & 0
\end{array}
\right]\quad \tau_3 \equiv \left[
\begin{array}{lr}
1 & 0\\
0 & -1
\end{array}
\right]
\end{eqnarray}
The photon $A_\mu(x)$ and the neutral weak vector boson $Z^0_\mu(x)$
are defined as the linear combinations
\begin{eqnarray}\label{defa}
\left[
\begin{array}{c}
Z^0_\mu(x)\\
A_\mu(x)
\end{array}
\right] \equiv
\left[
\begin{array}{lr}
\cos(\theta_W) & -\sin(\theta_W)\\
\sin(\theta_W) & \cos(\theta_W)
\end{array}
\right] \left[
\begin{array}{c}
W_\mu^3(x)\\
B_\mu(x)
\end{array}
\right]
\end{eqnarray}
where $\theta_W$ is the Weinberg (weak mixing) angle,
\begin{eqnarray}\label{defthetaw}
g \sin(\theta_W) = g' \cos(\theta_W) = e > 0
\end{eqnarray}
($e$ = electric charge of the proton). Combining \eqref{defb},
\eqref{defw3} and \eqref{defa}, we can therefore write
\begin{eqnarray}\label{aofm}
A_\mu(x) & = & \frac{\sin(\theta_W)}{2 g} \thinspace {\rm
Tr}\left[\tau_3
\thinspace {\bm M}_\mu(x)\right] \nonumber\\
&+& \frac{\cos(\theta_W)}{2 g'} \thinspace {\rm
Tr}\left[\tau_0\thinspace {\bm M}_\mu(x)\right]
\end{eqnarray}
Setting $Z^0_\mu(x) = W^1_\mu(x) = W^2_\mu(x) = 0$ and $A_\mu(x) =
1$, inverting \eqref{defa} and plugging the results into
\eqref{defm} yields ${\bm M}_\mu(x)$ for a normalized pure photon
state:
\begin{eqnarray}\label{defma}
{\bm M}^A_\mu = \left[
\begin{array}{lr}
2e & 0\\
0 & 0
\end{array}
\right]
\end{eqnarray}
(completely delocalized and therefore monochromatic; an envelope can
be imposed without consequence for the present argument). Now
consider the effect on ${\bm M}_\mu(x)$ of a global
SU(2)$\times$U(1) transformation
\begin{eqnarray}\label{defu}
{\bm U} \equiv \exp\left({\frac{i}{2} \thinspace \omega_0 \tau_0} +
{\frac{i}{2} \thinspace \omega_j \tau_j}\right)
\end{eqnarray}
with U(1) parameter $\omega_0$ and SU(2) parameters ${\bm \omega}
\equiv (\omega_1, \omega_2, \omega_3)$:
\begin{eqnarray}\label{mtrans}
{\bm M}_\mu(x) \rightarrow {\bm M}'_\mu(x) = {\bm U} {\bm
M}_\mu(x){\bm U}^\dagger
\end{eqnarray}
Substituting this into \eqref{aofm} and using the cyclic property of
the trace,
\begin{eqnarray}\label{atrans}
A_\mu(x) \rightarrow A'_\mu(x) & = & \frac{\sin(\theta_W)}{2 g}
\thinspace {\rm Tr}\left[{\bm U}^\dagger \thinspace \tau_3
\thinspace {\bm U}
{\bm M}_\mu(x)\right] \nonumber\\
&+& \frac{\cos(\theta_W)}{2 g'} \thinspace {\rm Tr}\left[{\bm
M}_\mu(x)\right]
\end{eqnarray}
U(1) transformations associated with $\omega_0\tau_0$ drop out.
Specializing to the pure photon state ${\bm M}^A_\mu$ of
\eqref{defma}, introducing the shorthand
\begin{eqnarray}
\omega_\bot^2 \equiv {\omega_1^2 + \omega_2^2}
\end{eqnarray}
\begin{eqnarray}
\omega \equiv \sqrt{\omega_\bot^2 + \omega^2_3}
\end{eqnarray}
and doing the traces, \eqref{atrans} reduces to
\begin{eqnarray}\label{pureatrans}
A_\mu & \rightarrow & A'_\mu = \ell({\bm \omega}) A_\mu
\end{eqnarray}
with
\begin{eqnarray}\label{defl}
\ell({\bm \omega}) \equiv \sin^2(\theta_W) \frac{\omega_\bot^2
\cos(\omega) + \omega_3^2}{\omega^2} + \cos^2(\theta_W)
\end{eqnarray}
(see Fig. \ref{fig:coeff_contour}, \ref{fig:coeff_persp}). Together
with \eqref{pureatrans}, equation \eqref{defl} describes the effect
of a global transformation with SU(2) parameters ${\bm \omega}$ (and
arbitrary U(1) parameter $\omega_0$) on a pure photon state. On its
own, since $A_\mu(x)$ is proportional to the photon number operator,
it gives us the fraction of photons surviving the transformation
(the rest having turned into $Z^0$s and linear combinations of
$W^1$s and $W^2$s, i.e. $W^\pm$s). The full import of this residual
luminosity $\ell({\bm \omega})$ will become evident in Section
\ref{section:distributions}.

\section{Electroweak phase transition}\label{section:ewpt}

In order to give the standard model fields (inertial) mass without
explicitly breaking the gauge symmetry of the Lagrangian (needed for
renormalizability \cite{thooft1971a}\cite{thooft1971b}), the weak
isospin doublet
\begin{eqnarray}\label{defphi}
\Phi(x) = \begin{bmatrix}
\phi^+(x)\\
\phi^0(x)
\end{bmatrix}
\end{eqnarray}
is introduced, where $\phi^+(x)$ and $\phi^0(x)$ are complex Lorentz
scalars transforming under SU(2)$\times$U(1) according to
\begin{eqnarray}\label{phitrans}
\Phi(x) \rightarrow \Phi'(x) = {\bm U} \Phi(x)
\end{eqnarray}
with the ${\bm U}$ of \eqref{defu}. Augmenting the Lagrangian with a
potential term featuring a manifold of degenerate ground states for
$\Phi(x) \ne 0$ (all connected by ${\bm U}$) turns $\Phi(x)$ into a
Higgs field \cite{higgs1964}\cite{higgs1966} which breaks the
symmetry dynamically by picking a non-zero VEV (vacuum expectation
value) $\langle\Phi\rangle$ in a specific ground state. The
asymmetric VEV splits the interaction terms between $\Phi(x)$ and
the other fields into effective mass terms and new interaction terms
involving $\Phi(x)$ excitations above the selected ground
state\footnote{For an alternative, perhaps more intuitive derivation
of $\ell({\bm \omega})$, diagonalize the gauge boson mass matrix for
arbitrary $\langle\Phi\rangle$ and take the scalar product of
massless eigenstates, i.e. photons, for VEVs related by the global
transformation parameter ${\bm \omega}$ \cite{anderberg2008}.}.

In the early universe, this electroweak phase transition (EWPT) is
expected to have occurred when the temperature of the expanding
plasma fell below the electroweak scale, $T_{EW} \approx
250\thinspace GeV \approx 2.9\cdot 10^{15}\thinspace K$. Plugging
$T_{EW}$ into the standard big bang model yields a cosmological
scale factor $\sim 10^{-15}$ and a Hubble radius $r_H \approx
2.9\cdot 10^{-23}$ relative to the present epoch, i.e. causally
connected patches roughly $1\thinspace cm$ across, each selecting a
VEV independently of the others.

Much effort has gone into studying the onset and early stages of the
EWPT, primarily due to their implications for baryogenesis
\cite{dine2004}\cite{trodden1998}. Scenarios motivated by the latter
generally begin with $O(1)$ bubbles of $\langle\Phi\rangle \ne 0$
nucleating within each causal patch and then expanding into the
surrounding plasma at some small fraction of the speed of light
(typically $v_w \sim 0.1$; but see \cite{moore2000} for much lower
estimates) determined by the equilibrium between outward pressure
and plasma friction. Colliding bubbles are believed to have
undergone repeated bounces and reheating before finally merging
\cite{kurki-suonio1996}, leading to a period of slow average growth
lasting several orders of magnitude longer than their initial
expansion. Such a (first order, i.e. bubbly) phase transition has
been ruled out for the minimal standard model
\cite{kajantie1996}\cite{csikor1999} but remains a viable
possibility in its various extensions.

In the minimal standard model, the EWPT is a continuous crossover,
i.e. an intrinsically non-perturbative -- and therefore analytically
challenging -- process. While one may argue that the natural speed
scale in this case is that of sound in a relativistic plasma, $v_s =
\sqrt{1/3}$, little is actually known with certainty about the
dynamics involved.

\section{Vacuum realignment}\label{section:realignment}

Fortunately, apart from an overall scale factor set by the amount of
time needed to convert all space to the broken symmetry phase, the
late epoch should be relatively unaffected by the early dynamics.
Once all space has been converted, the problem boils down to the
realignment of $\langle\Phi\rangle$ across adjacent causal patches,
no matter how they got their initial VEVs.

To avoid a common source of confusion, we need to clarify the
meaning of the term ``realignment''. In the absence of gauge
interactions, it is unambiguous: when $\Phi(x)$ varies across space,
there is gradient energy, which is minimized by $\Phi(x)$ evolving
to a constant $\langle\Phi\rangle$. With gauge interactions, as long
as $\Phi(x)$ does not go to zero anywhere in the volume under
consideration, it is always possible to substitute a
position-dependent ${\bm U(x)}$ into \eqref{phitrans} and ``gauge
away'' any misalignment in $\Phi(x)$. But such a local
transformation requires that we also use
\begin{eqnarray}\label{mgaugetrans}
{\bm M}_\mu(x) \rightarrow {\bm M}'_\mu(x) &=& -i\thinspace {\bm
U}(x)\left[\partial_\mu {\bm U}^\dagger(x) \right] \nonumber\\
&&+\thinspace {\bm U}(x) {\bm M}_\mu(x){\bm U}^\dagger(x)
\end{eqnarray}
in lieu of \eqref{mtrans}. What this amounts to is a change of
variables: we are trading gradients in $\Phi(x)$ for gradients in
${\bm M}_\mu(x)$ \footnote{Introductory QFT texts tend to focus on
infinitesimal transformations, sometimes leaving the full form
\eqref{mgaugetrans} to more advanced discussions of non-perturbative
solutions. Unfortunately, this practice seems to have spawned a
legion of phenomenologists who genuinely believe that the standard
electroweak model has only one physical vacuum, to which all other
vacua can be transformed without ulterior consequences. What they
make of sphalerons, the saddle point solutions connecting different
vacua which underpin the relevance of electroweak theory to
baryogenesis, is a mystery.}.

The obvious way to cure the resulting ambiguity is to choose a
(complete) gauge fixing condition and then stick with it. The
natural choice when considering multiple vacua (as opposed to doing
perturbation theory in one of them, the traditional business of
particle physics) is a physical gauge which leaves the internal
symmetry of the theory manifest, allowing the concept of alignment
to retain its intuitive meaning (unless otherwise stated, this point
of view will be implied for the rest of this paper). But the
ultimate arbiter is always energy. A true vacuum is a global energy
minimum of the theory; if the energy density within a given volume
is everywhere at such a minimum, we say that we have alignment
within that volume, even though our particular choice of gauge may
make individual field components look anything but aligned.

This brings us to another, closely related source of confusion: the
gauge trajectory argument. If moving along a spacetime trajectory
takes us through a sequence of $\Phi(x)$ values related by a gauge
trajectory, i.e. by a sequence of successive infinitesimal
transformations, then (and only then) using the inverse of this
gauge trajectory for ${\bm U(x)}$ in \eqref{phitrans} and
\eqref{mgaugetrans} will realign $\Phi(x)$ along the spacetime
trajectory without affecting the energy carried by the gauge fields.
The oft-quoted argument, due to Turok, that textures (knot-like,
gradient-only $\Phi(x)$ configurations) become true vacuum
configurations when $\Phi(x)$ has gauge interactions
\cite{turok1989} is a special case of this observation. It does not
imply, as it's sometimes misconstrued to do, that gradient-only
$\Phi(x)$ configurations residing entirely on the vacuum manifold
are trivial in gauge theories, only that they can generally be
expected to be unstable (Turok's use of the word ``become'' may be
partly to blame for this common misunderstanding; it should be read
as ``evolve to''). If we arrange $\Phi(x)$ in such a configuration
on a spacelike 3-surface with all gauge fields ${\bm M}_\mu(x)$ set
to zero, the result is indistinguishable from having the same
configuration in the corresponding ungauged theory, and so will
necessarily have positive energy. If we now let the field equations
run their course, we will see the energy being dispersed as ${\bm
M}_\mu(x)$ picks up. There is no question about this being a real,
physical process playing out over time. The only question is: how
much time?\footnote{Turok argued that the answer is a microphysical
time, set by the time scale of gauge interactions, but the argument
fails when massless gauge fields remain after SSB. See
\cite{anderberg2007} and Sections IV.A, IV.B in
\cite{anderberg2008}.}

In the case of a localized $\Phi(x)$ configuration living in an
otherwise empty vacuum (such as a texture), energy can disperse in
all directions at the speed of light; in practice the only relevant
time scale is that of the gauge interactions. The natural
expectation is then for the configuration's peak energy density to
decay exponentially with a half-life on the interaction scale (see
\cite{turok1990} for actual simulation results on electroweak
texture decay). In the more complicated case of a random
configuration filling all space, there is a second time scale: the
average propagation time to the first recurrence of the field values
(derivatives included), better known as inverse temperature. Once we
hit such a recurrence -- and in an infinite space, one is guaranteed
to occur to any desired degree of precision in every spatial
direction -- we have a periodic boundary condition, implying
conservation of energy, as opposed to the absorbing boundary
conditions which allow localized configurations living in an empty
vacuum to simply vanish from sight. The natural expectation is then
for energy density to settle into a semi-periodic pattern.

The weak spot in the gauge trajectory argument should now be
evident: the argument tells us that ${\bm M}_\mu(x)$ \emph{can} be
arranged so that a $\Phi(x)$ configuration which never leaves the
vacuum manifold may be gauged away ``for free'', but it does not
relieve us of having to explain \emph{where} the energy of the
initial field configuration ends up, nor \emph{how} it is carried
away. No explicit mechanism, no decay.

It is therefore necessary to consider the flow of conserved
quantities between $\Phi(x)$ and all other fields, fermions
included. In this paper, we are primarily concerned with $\Phi(x)$
configurations which can interact with photons, i.e. which carry
electric charge. A single, massive, charged electroweak boson -- the
ultimate localized state of the theory -- will of course decay on
the time scale of electroweak interactions, but that's beside the
point; the issue at hand is the lifetime of extended field
configurations on the vacuum manifold, not of isolated excitations
above it. To state the obvious, a boson VEV is not a ``dust cloud''
of distinct on-shell particles (if it were, the electroweak vacuum
itself, being a Higgs condensate, would decay as fast as a lone
Higgs boson); it's a continuous quantity whose evolution is
determined by the field equations derived from the theory's
effective action (classical action + radiative corrections).

Fortunately, we need not carry out the full program here. Thanks to
the work on pair production initiated a long time ago by Schwinger
\cite{schwinger1951}, we know that it can be viewed as a tunneling
process putting virtual fermion pairs on shell at the boson field's
expense. As long as a spatial volume subtended by the Compton
wavelength $1/m_e$ of the lightest charged fermion, the electron,
contains at least unit electric charge and gradient energy $\geq m_e
\approx 511\thinspace keV$ (plus the negligible rest mass of a
neutrino) there can be spontaneous emission of electron +
antineutrino pairs. When charge and gradient energy densities fall
below these thresholds, the tunneling rate becomes exponentially
suppressed. (The elementary insight that charge conservation limits
the decay rate of a charged boson condensate, making it quite
distinct from the decay rate of a single boson, can also be arrived
at by traditional statistical mechanics, as demonstrated to dramatic
effect in \cite{dolgov2005}). Further dissipation must then be
catalyzed by interactions with the environment.

In cosmology, interactions are strictly between field modes with
wavelength $\leq r_H$. Short of renouncing locality, superhorizon
modes are decoupled. No local interaction can therefore dissipate
conserved quantities carried by such modes. If interactions between
a field and the environment effectively shut down at some point in
time, field modes which were outside the horizon at that time are
still around today.

Consider a $\Phi(x)$ configuration looking to lose some positive
electric charge. The cheapest catalyst is a free electron. Turning
it into a $511\thinspace keV$ neutrino does not cost any energy, but
a spatial charge density $\gtrsim e\thinspace m_e^3$ is still
required (the matrix element is the same as for pair creation, we
have simply reversed an external momentum). Once charge density
falls below this threshold, it can no longer be dissipated away
effectively, and so neither can the $\Phi(x)$ configuration carrying
it. Since the $\phi^+(x)$ has unit electric charge, this implies a
threshold energy density $\sim m_H m_e^3 = m_H^4 (m_e/m_H)^3$,
corresponding to a temperature factor $\sim (m_e/m_H)^{3/4} \approx
10^{-5}$ relative to $T_{EW}$, i.e. $T \sim 10\thinspace MeV \approx
10^{11}\thinspace K$; the leptonic era. Relative to the present
epoch, the cosmological scale factor was then $\sim 10^{-10}$ and
the Hubble radius $\sim 10^{-15}$, or $\sim 10^3\thinspace km$. This
gives us the scale beyond which $\Phi(x)$ configurations should have
been safe from dissipation.

Incidentally, the subsequent redshifting of $10\thinspace MeV$ by
the scale factor $10^{-10}$ lands us right at today's dark energy
scale, $10^{-3}\thinspace eV$, suggesting that Higgs gradients may
provide the missing energy density required for $\Omega \approx 1$.

To state the obvious once more, $10^3\thinspace km$ is a macroscopic
distance separated by some 23 orders of magnitude from the
electroweak scale, the natural focus of works on the EWPT. There is
therefore no contradiction between the commonly expected fast
dissipation of configurations with characteristic sizes on the
electroweak scale, like Z strings (unless stabilized by plasma
effects, as argued by Nagasawa and Brandenberger
\cite{nagasawa2003}) and long-lived modes with wavelength
$\gtrsim \thinspace 10^3\thinspace km$.

Summing up, the misalignment in $\langle\Phi\rangle$ after the EWPT
is a physical reality which can not simply be ``gauged away''. As
the Hubble radius grows, regions establishing causal contact for the
first time after the transition have to realign according to the
field equations. This realignment can not proceed at a speed faster
than light's.

A final clarification has proved necessary:

In theories supporting nontrivial mappings between spacetime and
internal symmetry space, realignment sooner or later hits a stopping
point in the form of a topological defect, i.e. a
$\langle\Phi\rangle$ configuration which can not be realigned
throughout all space without leaving the vacuum manifold, at an
energy cost on the order of the symmetry breaking scale (Kibble
mechanism \cite{kibble1976}). Below this scale, such defects are
therefore classically stable. Their importance for cosmology has
been well understood for decades and has spawned a vast literature.
In particular, it has long been appreciated that domain walls, i.e.
topologically stable, two-dimensional defects passing through
$\langle\Phi\rangle = 0$, would be catastrophic, as each Hubble
volume would quickly become dominated by a massive, single wall
\cite{harvey1982}\cite{vachaspati1984}. Theories with domain wall
solutions are therefore ruled out by observation.

\textbf{The domain structure referred to in the title has nothing to
do with topologically stable domain walls.} The term ``domain
boundary'', as opposed to ``domain wall'', was adopted to help keep
this distinction in mind.

\section{Domain structure}\label{section:domainstructure}

Unlike the onset and early stages of the EWPT, its later stages have
attracted little attention. Since the standard electroweak model
features no topological defects and no (known) dynamically stable
solutions \footnote{A stable oscillating solution (and potential
non-exotic dark matter candidate) was found numerically after this
was written \cite{graham2006}.}, to the extent that realignment has
even been recognized as a real, physical process, it has simply been
assumed to proceed at the rate typical of electroweak interactions,
without any consequences worthy of notice. There is therefore little
in the way of past results to help us gain some insight into its
dynamics. Ultimately, settling the issue will come down to massive
numerical simulations. Until such studies are performed, the
analytical intractability of the highly non-linear electroweak field
equations forces us to rely on heuristics and on analogy with other
physical systems.

In Section \ref{section:realignment}, conservation of energy led us
to expect the emergence of a semi-periodic spatial pattern. A
further inroad to the problem is again provided by conservation of
electric charge (Q).

Fixing a global coordinate system in weak isospin space implies
fixing a definition for Q. By the standard conventions, the two
complex Higgs components $\phi^+(x)$ and $\phi^0(x)$ of
\eqref{defphi} carry unit and zero Q, respectively. While the
universe as a whole is assumed to be electrically neutral, the
random choice of $\langle\Phi\rangle$ at the EWPT will therefore
initially result in a random charge distribution.

Consider a volume with radius $R \gg r_H$. It starts off containing
some net Q. On average, this net charge will flow outward. Since the
initial distribution is random, the trajectory of a charged test
particle starting from the center of the volume is a
three-dimensional random walk, covering an average distance
\begin{eqnarray}\label{defrmean}
\langle r\rangle \propto \sqrt{t}
\end{eqnarray}
in time $t$ and becoming fractal-like (i.e. statistically scale
invariant) at late times. When $\langle r\rangle = R$, the
distribution has become a set of charged shells enclosing neutral
regions with average radius $R$. As they collide, oppositely charged
shells annihilate, their enclosed volumes merging; equally charged
shells press against each other. By comparison with compression of
randomly packed spheres, the resulting partitioning of space can be
expected to consist of irregular polyhedra with an average of
$\approx 13$ faces \cite{coxeter1958}\cite{coxeter1961}.

The upshot is that the current operator for a complex scalar field
is all derivative terms. By tracking Q flows, we are therefore
tracing out regions where $\langle\Phi\rangle$ is not constant.
These are our domain boundaries. As time goes by and domains merge,
average domain size grows, reducing the number of domain boundaries
and their total surface area.

What we have here are all the essential features of a problem well
known to materials scientists: local interactions causing the
formation of flat surfaces separating polyhedral domains, followed
by minimization of total surface area by successive domain merging.
Such a system is known as a \textbf{foam}, the problem of its
evolution as ``foam coarsening'' (or ``grain growth''). Its main
attraction is universality: the underlying interactions do not
matter as long as they provide the above features. Polyhedral
solutions are indeed commonplace in non-linear field theories and
have also been explicitly constructed in the standard electroweak
model\footnote{Better still, the low energy effective field theory
of the electroweak boson sector is a gauged non-linear sigma model
(NLSM), which is easily verified to admit the large number of
solutions known from the plain NLSM, including polyhedral ones
\cite{anderberg2007}\cite{anderberg2008}.} \cite{kleihaus2004}.

Foam coarsening is a tricky problem. von Neumann famously solved the
two-dimensional case on the spot upon hearing about it, equating the
area rate of change of a two-dimensional domain to the number of its
sides \cite{neumann1952}, but the three-dimensional case remains an
open challenge. Two universal rules, first described by Plateau in
his classic work on soap films \cite{plateau1873}, are known to
apply: along each edge, three faces of the constituent polyhedra
meet at angles of $2\pi/3$; at each vertex, four edges meet at the
tetrahedral angle $\arccos(-1/3)$. As a consequence, the average
number of faces $\langle f\rangle$ and the average number of edges
per face $\langle n_e \rangle$ can be shown to satisfy
\begin{eqnarray}
\langle f\rangle = \frac{12}{6 - \langle n_e \rangle}
\end{eqnarray}

\begin{figure}
\includegraphics[scale=0.55]{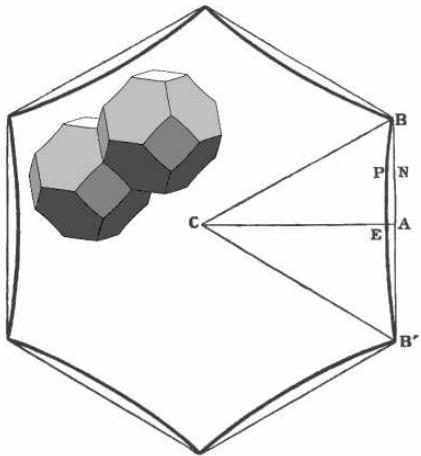}
\caption{\label{fig:kelvincombo}Two semi-regular tetrakaidecahedra
superimposed on an illustration of their section from Kelvin's paper
\emph{On the Division of Space with Minimum Partitional Area}
(1887). Real foams do not display such regularity, but their average
properties resemble those of the tetrakaidecahedron.}
\end{figure}

The fundamental difficulty in going beyond this level -- and the
crucial point of this section -- is that while the shaping of domain
boundaries is a local process proceeding on the time scale of the
underlying interaction, the foam's evolution, i.e. domain growth, is
not. When two domains merge, it is not an isolated event: their
nearest neighbors (and then \emph{their} nearest neighbors, and so
on) are also affected, ultimately forcing the whole foam to
rebalance. In the case at hand, currents will speed up, slow down or
even reverse, taking our charged test particle along on a continued
three-dimensional random walk. Since currents live on domain
boundaries, we are led to expect the statistical properties of the
test particle's trajectory, i.e. scale invariance and average radius
growth according to \eqref{defrmean}, to carry over to the foam
itself.

This does indeed turn out to be the case. After much experimental
and numerical work (not least on Potts models, believed to closely
approximate the finite temperature dynamics of SU(N) gauge theories
in general and of SU(2) in particular \cite{wipf2006}), a consensus
has emerged in recent years that the average volume growth rate of
three-dimensional domains with $f$ faces is well described by a
simple linear dependence on $f$ \cite{glazier2000}\cite{prause2000}
\begin{eqnarray}\label{defvgmean}
\langle V_f\rangle^{-1/3}\frac{d\langle V_f\rangle}{dt} = \kappa
\left(f - f_0\right)
\end{eqnarray}
whence
\begin{eqnarray}
f_0 \approx \frac{\langle f^2\rangle}{\langle f\rangle}
\end{eqnarray}
Empirically, $f_0 \approx 14 \pm 2$, well in line with our
guesstimate based on sphere crunching. Growth laws on the form
\eqref{defvgmean} (i.e. with right hand side depending only on the
number of faces, or more generally on topological features of the
domains) lead to the anticipated time dependence \eqref{defrmean}
for the average domain radius. This result is also expected on
dimensional grounds from the emergence of a \textbf{scaling state},
characterized by growing average domain size but time-independent
topological and area distributions \cite{mullins1989}, again
consistent with expectations from the random walk argument (scale
invariance).

The scaling state is a disordered one, not a regular structure
composed of individually near-optimal partitions such as Kelvin's
famous 14-faced tetrakaidecahedron (Fig. \ref{fig:kelvincombo}). The
discovery by Weaire and Phelan \cite{weaire1994} that an 8-cell
``repeat unit'' with $\langle f\rangle = 13.5$ and $\langle n_e
\rangle = 5.11$ achieves a more efficient (smaller total surface)
partitioning of a given volume than Kelvin's solution explains why:
a combination of many cells of different shape can actually have
lower total surface energy than a regular structure.

It bears emphasizing that these are universal results, depending
only on the assumption that the system strives for surface area
minimization, not on any details of the underlying interactions. In
equation \eqref{defvgmean}, all interaction dependence is
encapsulated in the constant $\kappa$. In particular, it can not be
underscored enough that the domain boundaries referred to here have
nothing in common with topologically stable defects: they do not
depend on non-trivial mappings between spacetime and internal
degrees of freedom for their existence, they have low energy
density, they give rise to a completely different domain structure
(the foam), and they positively must be unstable for domain growth
and emergence of the scaling state to be at all possible. Domains
grow by merging, i.e. by the disappearance of domain boundaries; if
the latter were stable, growth could not happen. The separation of
time scales and the decelerating evolution embodied in
\eqref{defrmean} are collective (and essentially geometric)
properties of the whole foam, not of individual domain boundaries.

\section{Number of domains}\label{section:domainnumber}

In a relativistic setting, \eqref{defrmean} must be qualified by the
requirement that the speed of light not be exceeded (locally). The
simplest {\it ansatz} satisfying this condition is
\begin{eqnarray}\label{defrelr}
\langle r(t)\rangle = \alpha \sqrt{1 + \beta^2 (t-t_\alpha)}
\end{eqnarray}
valid for $t >= t_\alpha$, where $t_\alpha$ marks the onset of the
scaling state,
\begin{eqnarray}\label{alpha}
\langle r(t_\alpha)\rangle = \alpha
\end{eqnarray}
with initial growth rate
\begin{eqnarray}\label{betaalphacondition}
\langle \dot r(t_\alpha)\rangle = \frac{\alpha \beta}{2} \le 1
\end{eqnarray}
and asymptotic growth rate
\begin{eqnarray}
\lim_{t \rightarrow \infty} \langle \dot r(t)\rangle = \frac{\alpha
\beta}{2 \thinspace \sqrt{t}}
\end{eqnarray}
Putting \eqref{defrelr} on the standard Robertson-Walker metric with
scale factor $a(t)$ yields the proper ensemble domain radius
\begin{eqnarray}\label{properradius}
\langle r_{RW}(t)\rangle = \frac{a(t)}{a(t_\alpha)} \thinspace
\langle r(t)\rangle + \int_{t_\alpha}^t d\tau \thinspace
\frac{a(\tau)}{a(t_\alpha)} \thinspace \langle \dot r(\tau)\rangle
\end{eqnarray}
with
\begin{eqnarray}
\langle \dot r(t)\rangle = \frac{\alpha \beta^2}{2 \sqrt{1 + \beta^2
(t-t_\alpha)}}
\end{eqnarray}

\begin{figure}
\includegraphics[scale=0.55]{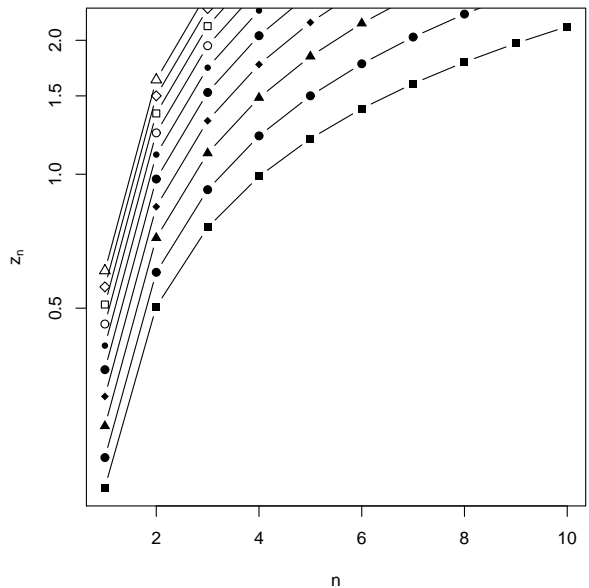}
\caption{\label{fig:redshifts}Expected domain boundary redshifts for
$\alpha/r_H(t_\alpha) \in [1,10]$, $\langle\dot r(t_\alpha)\rangle =
10^{-4}$, $t_\alpha = t(z_\alpha = 1089)$ (recombination) in a flat,
dust-matter Friedmann model.}
\end{figure}
Sweeping $\beta$ over the range $[0, 2/\alpha]$ will take us through
the possible values of $N_\Phi$, the average number of observable
$\langle\Phi\rangle$ domains in an arbitrary spatial direction. What
can we expect to find?

The limit case $\beta = 0$ is easily understood: it describes
domains of average size $\alpha$ being effectively ``frozen out'' at
$t = t_\alpha$ (by their growth rate becoming negligible compared to
that of the universe) and then simply coasting along with their
comoving volume. Assuming flatness, if $\alpha$ is the Hubble radius
at the time of the EWPT, the opposed effects of subsequent horizon
growth and metric expansion work out to $N_\Phi \sim 10^{23-15} =
10^8$. If a freeze-out occurs later, $N_\Phi$ can be substantially
lower. Section \ref{section:realignment} suggests $N_\Phi \lesssim
10^5$. One frozen domain per Hubble radius at recombination, $z
\approx 1089$ according to cosmic microwave background (CMB) data,
would translate to only $N_\Phi \sim 30$ today.

At the other end of the $\beta$ range, the Hubble radius poses an
absolute limit. As recently emphasized by Penrose \cite{penrose2004}
in a revival of the old homogeneity problem of pre-inflationary
cosmology, light reaching us now from quasars in opposite directions
must have originated in different electroweak domains, as those
sources have not been in causal contact since before the EWPT, which
occurred well after the end of inflation \footnote{A word of caution
is in place here. As it stands, Penrose's argument
\cite{penrose2004} can be interpreted as building on the notion that
the Weinberg angle $\theta_W$ was chosen randomly at the EWPT (see
pages 651 and 743), just like the direction of $\langle\Phi\rangle$.
While this may be the case in extended theories, it is not how the
standard electroweak model works. In the standard model, $\theta_W$
is just a parameter, not a field, and its value is the same
throughout spacetime.}. Thus,
\begin{eqnarray}\label{flatnrange}
1 \lesssim N_\Phi \lesssim 10^8
\end{eqnarray}
As will become evident in Section \ref{section:distributions}, too
small a number, i.e. a rate of realignment too close to the speed of
light, would indeed be incompatible with the observed isotropy of
the universe.

For a more detailed view, we must turn to equation
\eqref{properradius}. By isotropy, we should expect an infinite
sequence of domain boundaries at (average) proper distances
\begin{eqnarray}\label{properdistance}
d_n(t) = \langle r_{RW}(t)\rangle \thinspace (2n - 1) \quad\quad n
\in [1,\infty[
\end{eqnarray}
What we actually observe is light emitted at time $t_{em}$ and
reaching us at time $t$, so
\begin{eqnarray}\label{timeofemission}
\int_{t_{em}}^{t} \frac{d\tau}{a(\tau)} = d_n(t_{em})
\end{eqnarray}
Given $a(t)$, i.e. a cosmological model, we can solve
\eqref{timeofemission} for $t_{em}$ and plug the result into
\begin{eqnarray}
z = \frac{a(t)}{a(t_{em})} - 1
\end{eqnarray}
to obtain the observed redshift $z$. In practice, this must be done
numerically. See Appendix \ref{appendinx:positionscode} for easily
adaptable R \cite{r2005} code and Fig. \ref{fig:redshifts} for a
simple example (but not necessarily an unrealistic one, at least for
small $z$): a flat dust-matter model featuring $O(1)$ slow-growing
domains per Hubble radius at recombination.

To recap the story so far: while average domain radius is $\propto
\sqrt{t}$, the Hubble radius is $\propto t$ (assuming flatness).
Therefore, while any finite volume will eventually find itself
within a single domain, the number of observable domains grows with
time ($\propto \sqrt{t}$, again assuming flatness).

\section{Residual luminosity distributions}\label{section:distributions}

\begin{figure}
\includegraphics[scale=0.57]{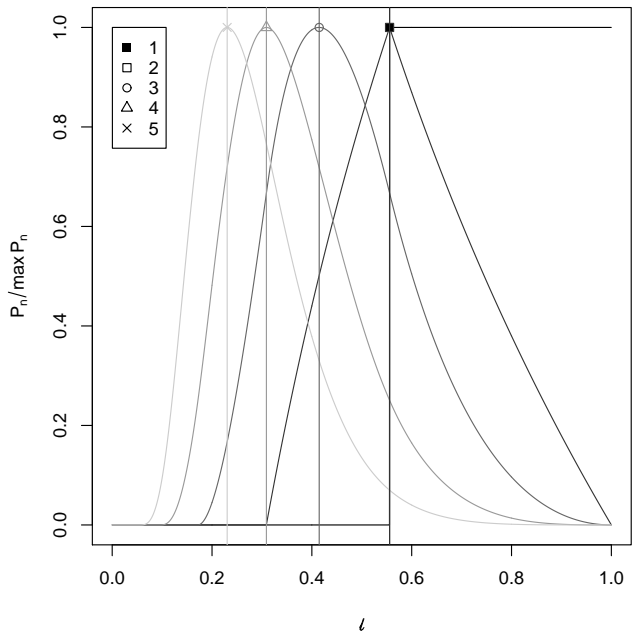}
\caption{\label{fig:pdf_lin_w_1_5}Probability distribution functions
$P_n(\ell)$ (vertically rescaled) for the residual luminosity $\ell$
after $n$ boundary crossings between $n + 1$ domains with
independent $\langle\Phi\rangle$.}
\end{figure}
Imagine a physicist in the $\langle\Phi\rangle$ domain $\mathfrak
D_1$ preparing a pure photon state ${\bm M}^A_\mu$ as defined by
\eqref{defma} and sending it to a colleague in the adjacent domain
$\mathfrak D_2$. To verify the purity of the received ${\bm M}_\mu$,
the colleague performs the operation \eqref{aofm} on it (e.g. by
measuring the peak interference amplitude with a locally produced
reference state). What will she see?

In their work, both physicists must implicitly rely on their
respective, local definition of a photon. The obvious way to
distinguish a photon from a $Z^0$ is that only the latter has mass,
a fact which is made explicit in the electroweak Lagrangian by a
change of basis in weak isospin space
\begin{eqnarray}\label{defunitary}
\langle\Phi\rangle \rightarrow \langle\Phi'\rangle = \begin{bmatrix}
0\\
\phi_0
\end{bmatrix}
\end{eqnarray}
making $\phi_0 \in \Re$ the only non-zero component of
$\langle\Phi\rangle$. This choice of basis aligns our coordinate
axes in weak isospin space with the direction which was picked by
the Higgs field at the EWPT. The SU(2)$\times$U(1) (sub)symmetry of
the ground state manifold guarantees the existence of a
transformation $\bm U$, as defined by \eqref{defu}, which achieves
this alignment. The Higgs and gauge fields in the new basis are
given by \eqref{phitrans} and \eqref{mtrans}, respectively. In this
basis, the $Z^0$ has a mass term and the photon does not.

But a global SU(2)$\times$U(1) transformation $\bm U$ can put
$\langle\Phi\rangle$ on the form \eqref{defunitary} in only one of
the two domains $\mathfrak D_1$ and $\mathfrak D_2$. In the other
domain, where $\langle\Phi\rangle$ is different, $\bm U$ has no
particular significance. Therefore, the local definition of a photon
is not the same in $\mathfrak D_1$ and in $\mathfrak
D_2$\footnote{This can be seen explicitly by diagonalizing the gauge
boson mass matrix for an arbitrary $\langle\Phi\rangle$
\cite{anderberg2007}}.

Formally, it is of course possible to perform a local transformation
and ``gauge away'' any misalignment in $\langle\Phi\rangle$
everywhere in the universe, but as we saw in Section
\ref{section:realignment}, only at the cost of using
\eqref{mgaugetrans} in lieu of \eqref{mtrans}, i.e. of trading
gradients in $\langle\Phi\rangle$ for gradients in ${\bm M}_\mu(x)$.
This amounts to a change of variables and of focus, from the Higgs
field inside domains to electrically charged gauge fields across
their boundaries. While it may facilitate the description of
processes at the boundaries, it does not affect the physics, and it
does not help our two physicists somewhere inside their respective
domain $\mathfrak D_1$ and $\mathfrak D_2$ to establish a common
frame of reference in weak isospin space. To that end, they need to
exchange photons and literally see what they get. If they choose to
analyze the experiment in terms of gauge fields across the boundary,
they must integrate along the entire path of the photons; if they
choose to think in terms of the Higgs in the bulk, they need only
consider the end points of the path.

We can now answer the question asked at the beginning of this
section: if the VEVs $\langle\Phi\rangle$ in $\mathfrak D_1$ and
$\mathfrak D_2$ are related by a transformation with U(1) parameter
$\omega_0$ and SU(2) parameters ${\bm \omega}$, a pure photon state
${\bm M}^A_\mu$ prepared in $\mathfrak D_1$ will be seen by an
observer in $\mathfrak D_2$ as a mix of $\ell({\bm \omega})$ parts
photons and $1 - \ell({\bm \omega})$ parts $Z^0$s and $W^\pm$s, with
$\ell({\bm \omega})$ given by \eqref{defl}. Physically, photons
crossing the boundary between two electroweak domains are partially
converted to weak vector bosons, which then quickly decay to
fermions and lower energy photons, leaving us with the residual
luminosity $\ell({\bm \omega})$.

This is the essence of the problem pointed out by Penrose
\cite{penrose2004}. If vacuum realignment proceeded at or near the
speed of light, we would have only O(1) domain boundaries within our
Hubble volume, causing glaring anisotropies. Unless we are prepared
to give up locality, the resolution of this apparent contradiction
between standard model and observation lies in the opposite
direction, i.e. in a subluminal rate of realignment allowing a
larger number of domains to even out such anisotropies. (More on
this in Section \ref{section:discussion}).

Since ${\bm \omega}$ was picked randomly at the EWPT, the value of
$\ell({\bm \omega})$ can not be predicted. The best we can hope for
is its probability distribution function (PDF). Viewing ${\bm
\omega}$ as the SU(2) parametrization
\begin{eqnarray}\label{unit3sphere}
[n_0, n_i] \to \cos(\omega/2) + i \tau_i \frac{\omega_i}{\omega}
\sin(\omega/2)
\end{eqnarray}
of the unit 3-sphere
\begin{eqnarray}\label{unit3sphere}
(n_0)^2 + (n_1)^2 + (n_2)^2 + (n_3)^2 = 1
\end{eqnarray}
(the Higgs vacuum manifold, up to a factor $|\langle\Phi\rangle|$)
we have
\begin{eqnarray}
\omega/2 &=& \arccos(n_0)  \label{oofn0} \\
\omega_i &=& \frac{\omega \thinspace n_i}{\sin(\omega/2)}
\label{oiofni}
\end{eqnarray}
Substituting Eqs. \eqref{oofn0}-\eqref{oiofni} into Eq. \eqref{defl}
then yields
\begin{eqnarray}
\ell({\bm n}) = 1 - 2 n_\bot^2 \sin^2(\theta_W)
\end{eqnarray}
i.e. luminosity is determined by
\begin{eqnarray}
n_\bot^2 = n_1^2 + n_2^2
\end{eqnarray}
The luminosity PDF therefore follows from that of $n_\bot^2$ over
the unit 3-sphere. For uniformly distributed ${\bm n}$, it is simply
a step function, $P_1(\ell) > 0$ for $\ell \in [1 - 2
\sin^2(\theta_W), 1]$.

While the dynamics may modify this result (because the electroweak
symmetry group is only a subgroup of the 3-sphere's O(4)) a uniform
${\bm n}$ distribution is the null hypothesis until numerical
simulation results become available\footnote{Originally, the PDF was
obtained by sampling an ${\bm \omega}$ grid, based on the assumption
that $\omega_1$, $\omega_2$ and $\omega_3$ are independent,
uniformly distributed stochastic variables $\in [0, 4\pi]$. This
skewed the result toward higher luminosity, $\langle\ell\rangle
\simeq 0.838$ vs. 0.778 for $\sin(\theta_W) \simeq 0.22216$, but did
not affect standard deviation much ($\sigma \simeq 0.133$ vs.
0.128).}.

Having obtained the PDF for one domain boundary crossing,
$P_1(\ell)$, we can easily extend our analysis to the distribution
for the residual luminosity $\ell = \ell_1 \cdot \ell_2 \cdot ...
\cdot \ell_n$ after $n$ crossings between $n + 1$ domains with
independent $\langle\Phi\rangle$. By Rohatgi's result for the
distribution of the product of two stochastic variables
\cite{rohatgi1976},
\begin{eqnarray}
P_2(\ell = \ell_1 \cdot \ell_2) = \int_{\ell}^{1}\thinspace dx
P_1(x) P_1(\ell/x) / x
\end{eqnarray}
and generally
\begin{eqnarray}
P_{n+1}(\ell = \ell_1 \cdot \ell_2 \cdot ... \cdot \ell_n) =
\int_{\ell}^{1}\thinspace dx P_1(x) P_n(\ell/x) / x
\end{eqnarray}
(The argument $\ell/x$ can be shifted around between the two PDFs in
the integrand by a trivial change of variables; for numerics, the
form shown is preferable, since it minimizes interpolation on the
grid at maximum gradient). R code for $P_n(\ell)$ is given in
Appendix \ref{appendinx:multicrosscode}.

\begin{figure}
\includegraphics[scale=0.57]{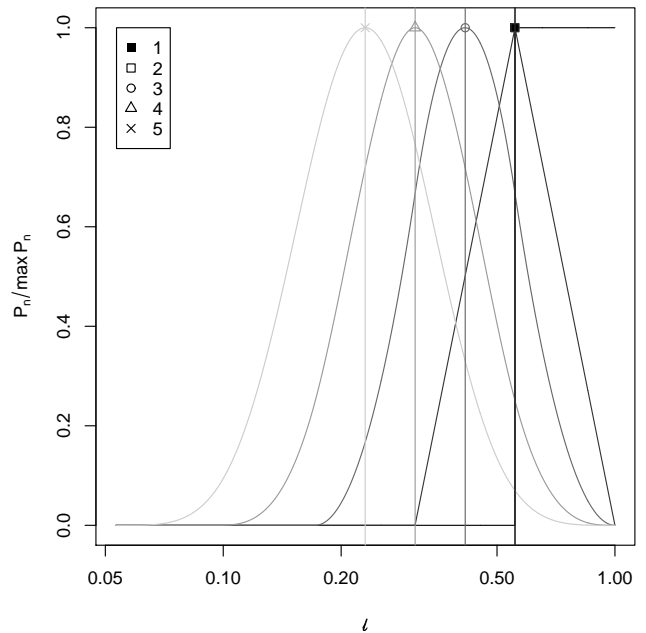}
\caption{\label{fig:pdf_log_w_1_5}Log-linear version of Fig.
\ref{fig:pdf_lin_w_1_5}. $P_n(\ell)$ is seen to quickly approach a
lognormal shape for $n > 3$.}
\end{figure}

\begin{figure}
\includegraphics[scale=0.57]{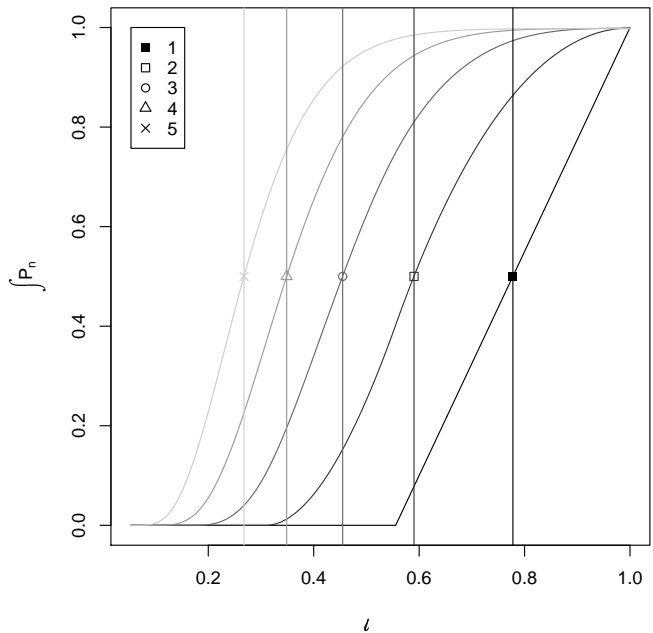}
\caption{\label{fig:cdf_lin_w_1_5}Cumulative probability
distribution functions $P_n(L \leq \ell)$ for the residual
luminosity $\ell$ after $n$ boundary crossings between $n + 1$
domains with independent $\langle\Phi\rangle$.}
\end{figure}

PDFs and cumulative distribution functions for $n \in [1,5]$ are
displayed in Fig. \ref{fig:pdf_lin_w_1_5}, \ref{fig:pdf_log_w_1_5}
and \ref{fig:cdf_lin_w_1_5}. Visually, their most striking feature
is the swiftness by which $P_n(\ell)$ morphs from a step function
for $n = 1$ to a lognormal for $n > 3$. But the real highlights of
this analysis are the first two records in Table
\ref{table:pdfstats}: they tell us to expect a residual luminosity
$\ell\ \sim 0.78 \pm 0.13$ for photons moving between adjacent
domains, and $\ell\ \sim 0.60 \pm 0.14$ for photons crossing two
domain boundaries, well in line with the $\ell/\ell_E$ ratio of
high-$z$ supernovae, $\sim 0.79$.

\begin{table}
\caption{\label{table:pdfstats}Mean, median, standard deviation,
skewness and excess kurtosis of $P_n(\ell)$ (5000 bins).}
\begin{ruledtabular}
\begin{tabular}{lrrrrr}
$n$&$\langle\ell\rangle$&$\tilde\ell$&$\sigma$&$\gamma_1$&$\gamma_2$\\
\hline\\
1& 0.7774 & 0.7780 & 0.1283 & 0.0082 & -2.970378\\
2& 0.6044 & 0.5904 & 0.4491 & 0.3564 & -2.949819\\
3& 0.4700 & 0.4550 & 0.1362 & 0.5845 & -2.944213\\
4& 0.3655 & 0.3488 & 0.1232 & 0.7681 & -2.946776\\
5& 0.2842 & 0.2678 & 0.1078 & 0.9285 & -2.953191
\end{tabular}
\end{ruledtabular}
\end{table}

\section{Supernova dimming}

The state of the art of $\ell/\ell_E$ determination is well
illustrated by \cite{riess2004}, where Riess et.al. quote an
intrinsic dispersion for individual supernovae (due to sparse
sampling and noisy photometry) of 0.15 magnitudes (in line with e.g.
\cite{schmidt1998} and \cite{filippenko2003}), and actual
dispersions about best fits (on a ``gold set'' of 157 SNe Ia
carefully selected from all available data) of 0.27 magnitudes for
$0.1 < z < 1.0$, increasing to 0.29 magnitudes for $z > 1.0$. By the
standard magnitude/luminosity relation
\begin{eqnarray}
l_2 / l_1 = 100^{(m_1 - m_2)/5}
\end{eqnarray}
this translates to an intrinsic luminosity uncertainty
$\pm100^{0.15/5} \sim \pm 0.15$, with actual $\ell/\ell_E$
dispersion ranging all the way from 0.78 to 1.28 for $0.1 < z <
1.0$, and from 0.77 to 1.31 for $z > 1.0$. From these figures,
making the (intentionally overoptimistic) assumption that all errors
are independent, the 1 $\sigma$ width for the combined $\ell/\ell_E$
ratio is $\epsilon \approx 0.25/\sqrt{157} \approx 0.02$, i.e.
\begin{eqnarray}\label{sne1aobserved}
\ell/\ell_E \approx 0.79 \pm 0.02
\end{eqnarray}

In terms of electroweak domains, the statistics is poorer. Under the
null hypothesis that our $\langle\Phi\rangle$ domain is an average
one, it has $\sim 14$ nearest neighbor boundaries (see Section
\ref{section:domainstructure}). With high-$z$ supernova searches
concentrated to the equatorial plane (so that the evolution of light
curves, needed for distance determination, may be tracked from
observatories on both hemispheres) we are probably looking through
fewer than half of them (see the section drawing in Fig.
\ref{fig:kelvincombo}). Based on Table \ref{table:pdfstats}, the 1
$\sigma$ expectation width for one boundary crossing is therefore
roughly $0.128/\sqrt{7} \approx 0.05$, i.e.
\begin{eqnarray}
\langle\ell_1\rangle/\ell_E \approx 0.78 \pm 0.05
\end{eqnarray}
This suggests that there is a single layer of domain boundaries
between us and high-$z$ supernovae\footnote{The higher
$\langle\ell\rangle$ of the original PDF left some room for multiple
boundary crossings, with a best fit including one- and two-boundary
crossings a $65/35$ mix. Assuming that an effective cutoff distance
$r_{cut}$ can be defined, such that observations are made in a
sphere of radius $r_{cut}$, this suggested an average distance to
the second domain boundary $\approx 0.65^{1/3} \thinspace r_{cut}
\approx 0.9 \thinspace r_{cut}$ and led to a very rough guesstimate
of $z \sim 1$.}.

The good match with one domain boundary allows a consistency check:
the boundary distance curves of Fig. \ref{fig:redshifts} illustrate
the fact that dimming should not start too close to home, lest we
end up expecting too many domain boundaries and too dark a universe
at higher redshifts. By inspection, the ``safe'' redshift is $z
\gtrsim 0.3$. Ideally (assuming that we are near the center of our
domain) we would like to see no sign of dimming at all for lower $z$
than this.

This condition is satisfied: in their 2003 analysis of published SNe
Ia data, Padmanabhan and Choudhury \cite{padmanabhan2003} found that
it split neatly in two subsets about $z = 0.25$, consistent with
different luminosity zero points and no continuous dimming at all --
exactly what crossing of a single domain boundary at $z = 0.25$
should look like. Their followup analysis with even more data
\cite{choudhury2005} showed that the split point can be moved all
the way to $z = 0.34$ with little consequence, definitely bringing
it into our ``safe'' zone. The conclusion that there is little or no
evidence of accelerated expansion for $z \lesssim 0.3$ was again
reached by Shapiro and Turner in \cite{shapiro2006}.

Summing up, the SNe Ia data appear to fit our expectations well,
based only on the standard electroweak and big bang models and
without any need for a cosmological constant.

\section{Discussion}\label{section:discussion}

This paper presents a novel picture of the universe as a foam-like
structure of large-scale electroweak domains in a scaling state,
arguing that this is both a natural consequence of the standard big
bang and electroweak models and a good match to SNe Ia observations
-- in fact the only one to date {\it sans} $\Lambda > 0$
\footnote{After completing the first version of this paper, the
author became aware of a dimming model which had been proposed
shortly before that version was written \cite{evslin2006}. In spite
of attractive phenomenological similarities -- a domain structure
causing dimming in discrete steps -- the underlying physical
motivation is quite different: \cite{evslin2006} invokes a stringy
braneworld scenario with compact extra dimensions to create domain
walls where photons mix with a hypothetical ``para-photon'', leaving
unresolved the problem of reconciling the required low energy
density of the walls with experimental limits on the string energy
scale, already in the hundreds of GeV.}.

In this picture, ``dark energy'' (in the restricted sense of an
unseen contribution to $\Omega$) is carried not by a hypothetical
new (set of) scalar field(s) with negative pressure, as in the
popular quintessence models, but rather by the long wavelength modes
of a very familiar one: the standard model Higgs (or the superset of
Higgs fields employed by standard model extensions). The SNe Ia
luminosity deficit is caused not by accelerated expansion, but by
photon conversion at domain boundaries. No exotic new physics is
needed. The most ``speculative'' assumption is that electroweak
domain growth is an unexceptional case of generic 3D domain growth.

In the reader's mind, this picture has likely already been met with
a strong objection (if not earlier, then upon review of Fig.
\ref{fig:redshifts}): what about the CMB? If we live inside an
irregular polyhedron with faces of different opacity, should we not
see huge anisotropies in the microwave background?

The first part of the answer is a counterquestion: what's a poor
photon to turn into? The $2.73\thinspace K$ of the present-day CMB
translate to $2.35\cdot 10^{-4}\thinspace eV$, well below the rest
mass of any known particle except the photon and -- perhaps -- the
lightest neutrino. With current neutrino oscillation data indicating
mass differences $\gtrsim 10^{-2}\thinspace eV$
\cite{melchiorri2005} (plus a much debated claim from the
Heidelberg-Moscow double beta decay experiment for $m_\nu >
0.17\thinspace eV$ \cite{klapdor2001}) conversion of CMB photons may
simply be kinematically forbidden, making electroweak domain
boundaries effectively transparent to them. In this regime, instead
of conversion, photons can be expected to undergo random scattering,
resulting in the diffuse glow seen in ordinary foams.

Two bounds on the lightest neutrino mass are thus implied: $m_\nu
\gtrsim 10^{-4}\thinspace eV$ to prevent readily visible CMB
anisotropies, $m_\nu \lesssim 1\thinspace eV$ for supernova dimming
without the reddening problem of dust models (the optical range
being roughly 2 to 3 $eV$). Incidentally, the lower bound also
explains an observed discrepancy in the relation between luminosity
and angular diameter distances for supernovae and radio sources
which has been used to argue against photon loss as an alternative
to Hubble acceleration \cite{bassett2004}: if domain boundaries are
transparent to microwaves, they are necessarily transparent to radio
waves, too.

It may be possible to improve the lower bound on $m_\nu$ by going
back in time, to higher CMB temperatures. If $m_\nu \lesssim
0.13\thinspace eV$, Heidelberg-Moscow notwithstanding, we will
eventually reach a point along the road to recombination at
$0.26\thinspace eV \sim z = 1089$ where CMB photons could be
converted to neutrinos (the ionization potential of hydrogen is
$13.6\thinspace eV$, but recombination occurs later due to photons
in the blackbody distribution tail and transitions between excited
states). Two considerations then come into play. First, the number
of domain boundaries cutting through the equatorial plane at a given
distance is roughly proportional to the number of domains out to
that distance. The WMAP angular resolution, $\approx 0.3^\circ$, is
enough to resolve 1200 individual faces, equivalent to $\sim 1200 /
6 = 200$ domains. Any $N_\Phi \gtrsim 200 \cdot 0.26/(2 m_\nu)$
should therefore be safe from direct observation by WMAP. Given
$m_\nu \lesssim 0.13\thinspace eV$, this implies $N_\Phi \gtrsim
200$. It then becomes a question of statistics: how large are the
anisotropies caused by differences between (individually
unresolvabe) faces?

Fluctuations in primary radiation which was exponentially damped by
$O(N_\Phi)$ boundary crossings are clearly irrelevant: $N_\Phi
\gtrsim 200$ implies a residual luminosity $\langle\ell\rangle
\lesssim 0.78^{200} \sim 10^{-22}$; with $N_\Phi = 10^5$, as
suggested by Section \ref{section:realignment}, $\langle\ell\rangle
\sim 10^{-10^4}$. Rather, the dominating contributions should be
from secondary photons which were redshifted below the $2 m_\nu$
threshold after only a few boundary crossings. Now, Fig.
\ref{fig:pdf_log_w_1_5} reminds us that residual luminosity
distributions become approximately lognormal, implying constant
relative standard deviation, already at $n = 3$. To estimate the
resulting anisotropies, we can therefore simply read off
$\sigma/\langle\ell\rangle \sim 10^{-1}$ from Table
\ref{table:pdfstats}, invoke the central limit theorem and divide by
the square root of the number of faces covered by a pixel at the $2
m_\nu$ threshold, $N_0 \approx (2 m_\nu / 0.26)\cdot N_\Phi$ domains
out.

For a sphere with radius $N_0$, the area of a cap subtended by angle
$\theta$ is $2\pi N_0^2 (1 - \cos(\theta/2))$. Tiling it with
tetrakaidecahedra of unit diameter gives us $7/(\pi/4)$ faces per
unit area, for a total of $28 N_0^2 (1 - \cos(\theta/2))$ faces.
This is actually an underestimate, since it does not take the uneven
spacing of domain boundaries into account; at large $z$, we should
rescale the radius by the comoving coordinate distance
\begin{eqnarray}\label{defDz}
D(z) = \frac{2\Omega z + (\Omega - 2)(\sqrt{1+\Omega z} -
1)}{\Omega^2 (1+z)}
\end{eqnarray}
(see e.g. \cite{weinberg1972}) to get
\begin{eqnarray}\label{defN}
N \approx 28\thinspace (N_0 D(z))^2\thinspace (1 - \cos(\theta/2))
\end{eqnarray}
(another way to see this is to first correct tile size for angular
diameter distance and then $N_0$ for scale factor). Inserting the
WMAP angular resolution $\theta = 0.3\cdot\pi/180$, $\Omega = 1$, $z
= 1089 \cdot 2 m_\nu / 0.26$, $m_\nu = 0.1\thinspace eV$ and $N_\Phi
= 10^5$ yields fluctuations $\sim 10^{-1}/\sqrt{N} \sim 10^{-5}$, in
line with observation. For smaller values, reduce $\Omega$
(everything else being equal, the old preferred $\Omega \approx 0.3$
buys a further factor 1/10) and/or increase $m_\nu$ and/or $N_\Phi$
(by equation \eqref{flatnrange} there is plenty of margin here too;
the order of magnitude argument pointing to $N_\Phi \sim 10^5$ is
certainly \emph{very} rough).

While this estimate is already in the right ballpark, keep in mind
that it ignores smoothing by random scattering (diffuse glow) and
the successive injection of high energy radiation from astrophysical
sources. Over time, down-conversion and scattering of this radiation
by domain boundaries should also build up a contribution to the CMB.
Intuitively, its spectrum should be the unit eigenstate of a Markov
chain through the decay channels available at each photon's energy,
with a peak about the average energy of photons produced by the last
decay in the chain; essentially the ``resonance peak'' of the
lightest particle capable of decaying to photons (yes, there is a
definite iconoclastic possibility lurking in this observation).
Computing it should not be hard, given a reliable Monte Carlo
generator of highly virtual weak boson decays (alas, apparently not
a priority in off-the-shelf simulation packages geared toward high
energy experiments).

A detailed CMB calculation would also require all neutrino masses to
be specified; the $\alpha$ and $\beta$ parameters of Section
\ref{section:domainnumber} to be extracted by numerical simulation
of the scaling state in the electroweak model under consideration,
in turn requiring knowledge of the Higgs mass(es); and of course a
consistent choice of cosmological expansion history. It would be a
large undertaking, but a systematic exploration of model+parameter
space might yield useful constraints on candidate cosmological
and/or extended electroweak models.

This resolution of the Penrose conundrum \cite{penrose2004} has a
problematic implication: the CMB is not quite the fossil we thought
it was. Long after it left the surface of ``last'' scattering, it
was interacting with domain boundaries, and later on it may have
picked up significant (dominating?) contributions from other
sources. This begs the question if it really is telling us so much
about the early universe as we currently like to believe. Its large
scale isotropy would undeniably be difficult to explain differently,
but at smaller angular separations, its features may be of more
recent origin. In particular, while none of this directly
contradicts inflation (even $\Omega \approx 1$ seems possible to
accomodate, at least at this early stage), it does call in question
the relevance of the CMB towards testing both inflationary and
alternative scenarios \cite{khoury2001}\cite{luminet2005}.

On the other hand, once the large-scale electroweak foam picture
starts sinking in, many new venues of investigation readily suggest
themselves. Do domain boundaries balance the energy budget of the
universe? What are their effects on structure formation? Can the
evidence for dark matter from cosmic shear be reinterpreted in terms
of domain boundary effects? Can they help explain observed CMB
anomalies \cite{magueijo2006}\cite{lieu2005}\cite{lieu2006}?

The answers are out there.

\appendix
\section{Domain boundary positions (R code)}
\label{appendinx:positionscode} {\footnotesize
\begin{verbatim}
# Replace flat dust with pet cosmos here
w <- 0
q <- 2 / (3 * (1 + w))
a <- function(t) { t^q }

# Auxiliary cosmological functions
arecip <- function(t) { 1/a(t) }
dprop <- function(t0, t1) {
   integrate(arecip, t0, t1,
   rel.tol=1E-13, abs.tol=1E-13)$value }
hR <- function(t) { a(t) * dprop(0, t) }

# Compute expected domain redshifts

redshifts <- function(
   a_alpha = 1/1090,  # a(t_alpha)
   r_alpha = 1,       # r(t_alpha)/rH(t_alpha)
   v_alpha = 0.00022, # v(t_alpha)
   maxZ = 5, maxN = 10)
{
   znVector <- array(NaN, c(1, maxN))

   # Domain functions
   r <- function(t) {
      alpha * sqrt(1 + beta2*(t - t_alpha)) }
   integrand <- function(t) {
      0.5 * alpha * beta2 * a(t) /
      sqrt(1 + beta2*(t - t_alpha)) }
   rprop <- function(t)
   {
      integral <- integrate(integrand,
                            t_alpha, t,
                            rel.tol=1E-13,
                            abs.tol=1E-13)
      (a(t) * r(t) + integral$value ) / a_alpha
   }
   dn <- function(t, n) { (2*n - 1) * rprop(t) }

   # Derived parameters
   a_alpha_dev <- function(t_alpha) {
      (a_alpha - a(t_alpha))^2 }
   t_alpha <- optimize(a_alpha_dev,
                       c(0, 1), tol=1E-30)$minimum

   alpha <- hR(t_alpha) * r_alpha
   beta <- 2 * v_alpha / alpha
   beta2 <- beta*beta

   # Main loop

   t_em_dev <- function(t, n) {
      (dprop(t, 1) - dn(t, n))^2 }

   for(n in 1:maxN)
   {
      t_em <- optimize(t_em_dev,
                       c(0, 1), n = n, tol=1E-30)
      z <- a(1)/a(t_em$minimum) - 1
      znVector[n] <- z

      if ((t_em$minimum < t_alpha) ||
          (z > maxZ)) break
   }

   znVector
}

# Main program
redshifts()
\end{verbatim}
}

\section{Luminosity distribution  for n = 1 (R code)}
\label{appendinx:onecrosscode} {\footnotesize
\begin{verbatim}
sw2 <- 0.22216              # sin^2(theta_W)

oneCrossLuminosity <- function(n)
# Return PDF and CDF of luminosity
# Simple step function for input to nCrossLuminosity()
 {
   pdf <- array(0, c(n))
   cdf <- array(0, c(n))

   i <- round(n*(1.0 - 2.0*sw2))

   pdf[i + 1:(n - i)] <- 1.0/(n - i + 1);

   cdf[1] <- pdf[1]
   for (i in 2:n) { cdf[i] <- pdf[i] + cdf[i-1] }

   # Return results
   list(pdf=pdf, cdf=cdf)
}

# Main program
sample <- sampleLuminosity(5000)
save(sample, file="sample.txt", ascii=TRUE)
\end{verbatim}
}

\section{Luminosity distributions for n $\geq$ 1 (R code)}
\label{appendinx:multicrosscode} {\footnotesize
\begin{verbatim}
nCrossLuminosity <- function(maxCross, n1pdf, n1cdf)
# Compute PDF&CDF for 2 to maxCross boundary crossings
{
   n <- length(n1pdf)

   pdf <- array(0, dim=c(maxCross, n))
   cdf <- array(0, dim=c(maxCross, n))

   for (i in 1:n)
   {
      pdf[1, i] <- n1pdf[[i]]
      cdf[1, i] <- n1cdf[[i]]
   }

   if (maxCross > 1)
   {
      for (nCross in 1:(maxCross - 1))
      {
         cumulative <- 0
         for (iy in 1:n)
         {
            integral <- pdf[nCross, n] * n1pdf[iy] / iy

            if (iy < n)
            {
               for (ix in (iy+1):n)
               {
                  integral <-
                     integral +
                     pdf[nCross, (n*iy) %/% ix] *
                     n1pdf[ix]/ix
               }
            }

            integral <- n*integral
            pdf[nCross + 1, iy] <- integral
            cumulative <- cumulative + integral
            cdf[nCross + 1, iy] <- cumulative
         }
      }
   }

   # Return PDFs and CDFs

   list(pdf=pdf, cdf=cdf)
}

# Main program
multi <- nCrossLuminosity(5, sample$pdf, sample$cdf)
save(multi, file="multi.txt", ascii=TRUE)
\end{verbatim}
}

\begin{table}
\caption{\label{table:cdfparams}Parameters of fit and error squared
$\epsilon^2$ for CDF parametrization $F(\ell; a, b, c, d)$
\eqref{cdffit}.}
\begin{ruledtabular}
\begin{tabular}{lrrrrr}
$n$&$a$&$b$&$c$&$d$&$\epsilon^2$\\
\hline\\
1 & 0.29650820 & 0.4238114 & -0.1885901 & 0.05361303 & 6.011231\\
2 & 0.05948206 & 0.7159291 & -0.3665269 & 0.1868697  & -0.03652571\\
3 & 0.09153357 & -2.05914  & 72.6966    & 14.41858   & -0.3107125\\
4 & 0.01036675 & 0.3067824 & -0.3199721 & 0.1081582  & -0.08748841\\
5 & 0.4831496  & -2.23015  & 31.56453   & 22.18672   & -0.05052748
\end{tabular}
\end{ruledtabular}
\end{table}

\begin{table*}
\caption{\label{table:pdfparams}Parameters of fit and error squared
$\epsilon^2$ for PDF parametrization $g(\ell; a, b, c, d, e)$
\eqref{pdffit}.}
\begin{ruledtabular}
\begin{tabular}{lrrrrrr}
$n$&$a$&$b$&$c$&$d$&$e$&$\epsilon^2$\\
\hline\\
1 & 2.484335e-05 & -0.1780371 & 0.2699788 & 5.571233e-08 & 0.0002125698 & 8.168009e-05\\
2 & 1.198534e-06 & -0.5352473 & 0.2612992 & -1.576451e-07 &
0.0001519190 & 9.103221e-05\\
3 & 1.897927e-07 & -0.6941752 & 0.332284 & -2.835761e-08 &
0.0003936519 &
-0.0003655848\\
4 & 8.499097e-08 & -0.9647373 & 0.3736279 & -1.269617e-08 &
0.0003422317 & -0.0003472561\\
5 & 5.919239e-08 & -1.235350 & 0.4105253 & -1.767185e-08 &
0.0003091985 & -0.0003447646
\end{tabular}
\end{ruledtabular}
\end{table*}

\section{Distribution fits}

\label{appendinx:parameterizations} { For $n > 1$, the CDFs $P_n(L
\leq \ell)$ of Fig. \ref{fig:cdf_lin_w_1_5} are found to be well
parameterized by
\begin{eqnarray}\label{cdffit}
F(\ell; a, b, c, d) = \frac{F_{ln}(\ell^a; b, c) \thinspace
\exp(d\thinspace \ell^a)}{F_{ln}(1; b, c) \thinspace \exp(d)}
\end{eqnarray}
where $F_{\ln}(x; b, c)$ is the lognormal CDF
\begin{eqnarray}
F_{\ln}(x; b, c) \equiv \frac{1}{2}\left\{1 + {\rm
erf}\left(\frac{\ln(x - b)}{c \sqrt{2}}\right)\right\}
\end{eqnarray}
${\rm erf}(x)$ is the error function
\begin{eqnarray}
{\rm erf}(x) \equiv \frac{2}{\sqrt{\pi}} \int_{0}^{x}dy \thinspace
\exp(-y^2)
\end{eqnarray}
and the parameters $a, b, c, d$ (plus total error squared
$\epsilon^2$) obtained on a 5000$\times$5000 point grid are given in
Table \ref{table:cdfparams}.

Differentiating \eqref{cdffit} in $\ell$ yields the corresponding
parametrization of $P_n(\ell)$:
\begin{eqnarray}
f(\ell; a, b, c, d) \equiv \frac{d}{dx} F(\ell; a, b, c, d) =\nonumber\\
\frac{\left(f_{ln}(\ell^a; b, c) + d \thinspace F_{ln}(\ell^a; b,
c)\right) \thinspace \exp(d\thinspace \ell^a) \thinspace a
x^{a-1}}{F_{ln}(1; b, c) \thinspace \exp(d)}
\end{eqnarray}
where $f_{ln}(x; b, c)$ is the lognormal PDF,
\begin{eqnarray}
f_{ln}(x; b, c) \equiv \frac{1}{2 c x\sqrt{2\pi}}
\exp\left({-\frac{\left(\ln{x} - b\right)^2}{2 c^2}}\right)
\end{eqnarray}

For the first couple of $n$, fitting directly on $P_n(\ell)$ with
\begin{eqnarray}\label{pdffit}
g(\ell; a, b, c, d, e) \equiv f_{ln}(\ell; a, b) \left\{ \frac{c}{1
- x} + d + e x \right\}
\end{eqnarray}
achieves a better reproduction of the sharp peak dominating the PDF.
Parameters obtained on a 5000$\times$5000 point grid are given in
Table \ref{table:pdfparams}. }

\begin{acknowledgments}
I wish to thank Roger Penrose for making the remarks
\cite{penrose2004} which inspired me to write this paper, Eliana
Vianello for her constant encouragement, GianCarlo De Pol for
illuminating questions and everybody else whose feedback helped
clarify where more detail was needed to prevent confusion.

This work is dedicated to my parents.
\end{acknowledgments}

\bibliography{basename of .bib file}

\begin{thebibliography}{kurki-suonio1996}
\bibitem{riess1998}A.G. Riess et al (1998) Astron. J. 116, 1009.
\bibitem{perlmutter1998}S. Perlmutter et al (1998) Nature (London) 391, 51-54.
\bibitem{perlmutter1999}S. Perlmutter et al (1999) Astrophys. J. 517, 565.
\bibitem{schmidt1998}B. Schmidt et al (1998) Astrophys. J. 507, 46-63.
\bibitem{filippenko2003}A. Filippenko (2003) in Carnegie Observatories
Astrophysics Series, Vol. 2: \emph{Measuring and Modeling the
Universe}, ed. W. L. Freedman (Cambridge University Press,
Cambridge) (\url{http://arxiv.org/astro-ph/0307139}).
\bibitem{riess2004}A.G. Riess et.al. (2004) Astrophys. J. 607, 665-687
(\url{http://arxiv.org/abs/astro-ph/0402512}).
\bibitem{schaefer2006}B.E. Schaefer (2006) Astrophys. J., in press
(\url{http://arxiv.org/abs/astro-ph/0612285}).
\bibitem{blanchard2003}A. Blanchard, M. Douspis, M. Rowan-Robinson,
S. Sarkar (2003) Astron. Astrophys. 412, 35-44
(\url{http://arxiv.org/abs/astro-ph/0304237}).
\bibitem{bonanos2006}A.Z. Bonanos et al (2006) Astrophys. J. 652,
313 (\url{http://arxiv.org/abs/astro-ph/0606279}).
\bibitem{wang2006}L. Wang, D. Baade, F. Patat (2006) Science, in
press (\url{http://arxiv.org/abs/astro-ph/0611902}).
\bibitem{glashow1961}S.L. Glashow (1961) Nucl. Phys. 22, 579.
\bibitem{weinberg1967}S. Weinberg (1967) Phys. Rev. Lett. 19, 1264.
\bibitem{salam1968}A. Salam (1968) in \emph{Elementary Particle Physics}
(Nobel Symp No. 8), ed. N. Svartholm (Almqvist and Wiksell,
Stockholm).
\bibitem{thooft1971a}G. 't Hooft (1971) Nucl. Phys. B33, 173.
\bibitem{thooft1971b}G. 't Hooft (1971) Nucl. Phys. B35, 167.
\bibitem{higgs1964}P.W. Higgs (1964) Phys. Rev. Lett. 13, 509.
\bibitem{higgs1966}P.W. Higgs (1966) Phys. Rev. 145, 1156.
\bibitem{anderberg2008}T. Anderberg (2008) arxiv:0804.2284
(\url{http://arxiv.org/abs/0804.2284}).
\bibitem{dine2004}M. Dine, A. Kusenko (2004) Rev. Mod. Phys. 76, 1
(\url{http://arxiv.org/abs/hep-ph/0303065}).
\bibitem{trodden1998}M. Trodden (1999) Rev. Mod. Phys. 71,
1463-1500 (\url{http://arxiv.org/abs/hep-ph/9803479}).
\bibitem{moore2000}G.D. Moore (2000) JHEP 0003, 006
(\url{http://arxiv.org/abs/hep-ph/0001274}).
\bibitem{kurki-suonio1996}H. Kurki-Suonio, M. Laine (1996) Phys. Rev. Lett. 77,
3951-3954 (\url{http://arxiv.org/abs/hep-ph/9607382}).
\bibitem{kajantie1996}K. Kajantie, M. Laine, K. Rummukainen, M. Shaposhnikov (1996) Phys. Rev. Lett. 77, 2887-2890
(\url{http://arxiv.org/abs/hep-ph/9605288}).
\bibitem{csikor1999}F. Csikor, Z. Fodor, J. Heitger (1999) Phys. Rev. Lett. 82, 21-24
(\url{http://arxiv.org/abs/hep-ph/9809291}).
\bibitem{turok1989}N. Turok (1989) Phys. Rev. Lett. 63, 2625.
\bibitem{anderberg2007}T. Anderberg (2007) arxiv:0711.3187
(\url{http://arxiv.org/abs/0711.3187}).
\bibitem{turok1990}N. Turok, J. Zadrozny (1990) Phys. Rev. Lett. 65, 2331.
\bibitem{schwinger1951}J. Schwinger (1951), Phys. Rev. 82, 664.
\bibitem{dolgov2005}A.D. Dolgov, F.R. Urban (2005) Astropart. Phys. 24, 289-300
(\url{http://arxiv.org/abs/hep-ph/0505255}).
\bibitem{nagasawa2003}M. Nagasawa, R. Brandenberger (2003) Phys.
Rev. D67, 043504 (\url{http://arxiv.org/abs/hep-ph/0207246}).
\bibitem{kibble1976}T.W.B. Kibble (1976) J. Phys. A9, 1387.
\bibitem{harvey1982}J.A. Harvey, E.W. Kolb, D.B. Reiss, S. Wolfram (1982) Nucl. Phys.
B201, 16.
\bibitem{vachaspati1984}T. Vachaspati, A. Vilenkin (1984) Phys. Rev. D30, 2036.
\bibitem{graham2006}N. Graham (2007) Phys. Rev. Lett. 98, 101801; ibid.
189904(E) (\url{http://arxiv.org/abs/hep-th/0610267}).
\bibitem{coxeter1958}H.S.M. Coxeter (1958) Illinois J. Math. 2, 746-758.
\bibitem{coxeter1961}H.S.M. Coxeter (1961) in \emph{Introduction to Geometry,
2nd ed.}, 405-411 (Wiley, New York).
\bibitem{kleihaus2004}B. Kleihaus, J. Kunz, K. Myklevoll (2004) Phys. Lett. B582,
187-195 (\url{http://arxiv.org/abs/hep-th/0310300}).
\bibitem{neumann1952}J. von Neumann (1952) \emph{Metal Interfaces}, 108 (American Society for Metals,
Cleveland).
\bibitem{plateau1873}J.A.F. Plateau (1873) \emph{Statique Experim\'{e}ntal et Th\'{e}orique des Liquides Soumis aux Seules
Forces Mol\'{e}culaires} (Gauthier-Villars, Paris).
\bibitem{wipf2006}A. Wipf et.al. (2006) in ``O'Raifeartaigh Symposium on Non-Perturbative and Symmetry Methods in Field
Theory'' (proceedings) (\url{http://arxiv.org/abs/hep-lat/0610043}).
\bibitem{glazier2000}J.A. Glazier, B. Prause (2000) in Foams, Emulsions and their Applications:
\emph{Current Status of Three-Dimensional Growth Laws}, 120-127,
ed. P. Zitha, J. Banhart, G. Verbist (Verlag MIT Publishing, Bremen)
(\url{http://biocomplexity.indiana.edu/jglazier/papers.php?action=view&cat=02a&id=60}).
\bibitem{prause2000}B.A. Prause (2000) \emph{Magnetic Resonance Imaging of
Structure and Coarsening in Three-Dimensional Foams} (Ph.D. thesis,
University of Notre Dame)
(\url{http://biocomplexity.indiana.edu/jglazier/papers.php?action=view&cat=02a&id=81}).
\bibitem{mullins1989}W.W. Mullins, J. Vi\~{n}als (1989) Acta metall. 37, 991.
\bibitem{weaire1994}D. Weaire, R. Phelan (1994) Phil. Mag. Lett. 69, 107-110.
\bibitem{penrose2004}R. Penrose (2004) \emph{The Road to Reality}, 744-746 (Knopf, New York).
\bibitem{r2005}R Development Core Team (2005) \emph{R: A language and environment for statistical computing} (Vienna, ISBN 3-900051-07-0)
(\url{http://www.R-project.org}).
\bibitem{rohatgi1976}V.K. Rohatgi (1976) \emph{An Introduction to Probability Theory and Mathematical Statistics} (Wiley, New York).
\bibitem{padmanabhan2003}T. Padmanabhan, T.R. Choudhury (2003), MNRAS, 344, 823.
\bibitem{choudhury2005}T.R. Choudhury, T. Padmanabhan (2005) Astron. Astrophys. 429, 807
(\url{http://arxiv.org/abs/astro-ph/0311622}).
\bibitem{shapiro2006}C. Shapiro, M.S. Turner, Astrophys. J. 649 (2006) 563
(\url{http://arxiv.org/abs/astro-ph/0512586}).
\bibitem{melchiorri2005}A. Melchiorri et.al. (2005) Nucl. Phys. B Suppl., Vol. 145, 290-294
(\url{http://arxiv.org/abs/astro-ph/0501531}).
\bibitem{klapdor2001}H.V. Klapdor-Kleingrothaus et.al. (2001) Mod. Phys. Lett. A 16, 2409.
\bibitem{bassett2004}B.A. Bassett, M. Kunz (2004) Astrophys. J. 607, 661-664
(\url{http://arxiv.org/abs/astro-ph/0311495}).
\bibitem{evslin2006}J. Evslin, M. Fairbairn (2006) JCAP 0602, 011
(\url{http://arxiv.org/abs/hep-ph/0507020}).
\bibitem{weinberg1972}S. Weinberg (1972) \emph{Gravitation and Cosmology}, 485 (Wiley, New York).
\bibitem{khoury2001}J. Khoury, B.A. Ovrut, P.J. Steinhardt, N. Turok (2001)
Phys. Rev. D64, 123522
(\url{http://arxiv.org/abs/hep-th/0103239}).
\bibitem{luminet2005}J.P. Luminet (2005) Phys. World 18, 22-28
(\url{http://arxiv.org/abs/physics/0509171}).
\bibitem{magueijo2006}J. Magueijo, R.D. Sorkin, astro-ph/0604410
(\url{http://arxiv.org/abs/astro-ph/0604410}).
\bibitem{lieu2005}R. Lieu, J.P.D. Mittaz (2005) Astrophys. J. 628, 583-593
(\url{http://arxiv.org/abs/astro-ph/0412276}).
\bibitem{lieu2006}R. Lieu, J.P.D. Mittaz, S.-N. Zhang (2006) Astrophys. J. 648,
176 (\url{http://arxiv.org/abs/astro-ph/0510160}).
\end{thebibliography}

\end{document}